# Electrical transport properties driven by unique bonding configuration in γ-GeSe

*Jeongsu Jang[1,†], Joonho Kim[1,†], Dongchul Sung[2,†], Jong Hyuk Kim[1], Joong-Eon Jung[1], Sol Lee[1,3], Jinsub Park[1], Chaewoon Lee[1], Heesun Bae[1], Seongil Im[1], Kibog Park[4,5], Young Jai Choi[1,\*], Suklyun Hong[2,\*], and Kwanpyo Kim[1,3,\*]*

[1]Department of Physics, Yonsei University, Seoul 03722, Korea.

[2]Department of Physics, Graphene Research Institute and GRI-TPC International Research Center, Sejong University, Seoul 05006, Korea.

[3]Center for Nanomedicine, Institute for Basic Science (IBS), Seoul 03722, Korea.

[4]Department of Physics, Ulsan National Institute of Science and Technology (UNIST), Ulsan, 44919, Korea.

[5]Department of Electrical Engineering, Ulsan National Institute of Science and Technology (UNIST), Ulsan, 44919, Korea.

[†]These authors contributed equally.

*Address correspondence to Y.J.C.(phylove@yonsei.ac.kr) S.H.(hong@sejong.ac.kr), K.K. (kpkim@yonsei.ac.kr)

**Abstract**

Group-IV monochalcogenides have recently shown great potential for their thermoelectric, ferroelectric, and other intriguing properties. The electrical properties of group-IV monochalcogenides exhibit a strong dependence on the chalcogen type. For example, GeTe exhibits high doping concentration, whereas S/Se-based chalcogenides are semiconductors with sizable bandgaps. Here, we investigate the electrical and thermoelectric properties of γ-GeSe, a recently identified polymorph of GeSe. γ-GeSe exhibits high electrical conductivity (~$10^6$ S/m) and a relatively low Seebeck coefficient (9.4 µV/K at room temperature) owing to its high *p*-doping level ($5 \times 10^{21}$ cm$^{-3}$), which is in stark contrast to other known GeSe polymorphs. Elemental analysis and first-principles calculations confirm that the abundant formation of Ge vacancies leads to the high *p*-doping concentration. The magnetoresistance measurements also reveal weak-antilocalization because of spin-orbit coupling in the crystal. Our results demonstrate that γ-GeSe is a unique polymorph in which the modified local bonding configuration leads to substantially different physical properties.

Group-IV monochalcogenides (MX, M = Ge, Sn; X = S, Se, Te) are a class of materials that have recently attracted attention in research and various applications.[1-3] This family of materials is a promising candidate for applications, such as energy conversion devices and phase change memory,[4,5] owing to their intriguing electronic, thermoelectric,[6-10] and ferroelectric properties,[11-15] as well as their unique bonding mechanisms.[16,17] The physical properties of group-IV monochalcogenides show a strong correlation with the types of chalcogen atoms (S, Se, and Te). For example, Ge and Sn monochalcogenides with S and Se exhibit semiconducting properties, whereas monochalcogenides with Te have small band gap and exhibit high doping concentrations.[6,18-20] Furthermore, alloying of different chalcogenides has also been pursued to control the physical properties of this material system.[10,21,22]

In theory, group-IV monochalcogenides are expected to have various polymorphs owing to the versatile bonding configurations in this family of materials.[23-28] Despite the theories, there has been a lack of experiments involving various polymorphic configurations for GeS and GeTe in which only one type of polymorph is known to exist at room temperature. Among Ge-based monochalcogenides, GeSe serves as a unique chalcogenide with various polymorphic configurations; three types of polymorphs (α-GeSe, β-GeSe, and γ-GeSe) have been successfully synthesized (**Figure 1**). Conventional α-GeSe has a layered puckered structure similar to black phosphorus (BP), which has strong in-plane anisotropic properties. The anisotropic structure of α-GeSe makes it an attractive material for electronics and optoelectronics.[29] The synthesis of β-GeSe, which shares a similar local bonding configuration with that of α-GeSe, has also been reported.[24] Recently, the synthesis of a new hexagonal polymorph of GeSe, γ-GeSe, has been demonstrated.[30] Although recent theoretical studies have provided results on the interesting electrical, ferroelectric, and thermoelectric properties of γ-GeSe,[31-33] these results are yet to be confirmed experimentally.

Herein, we report the electrical and thermoelectric properties of γ-GeSe. The temperature-dependent electrical conductivity of γ-GeSe confirms its metallic behavior owing to a high level of *p*-doping (~ 5×10$^{21}$ cm$^{-3}$). The value and temperature dependence of the measured Seebeck coefficient were also consistent with the high *p*-doping. Using experimental elemental analysis and first-principles calculations, we showed that the high doping is originated from the easy formation of Ge vacancies in the new polymorphic phase. Using magneto-transport measurements, we also confirmed the existence of weak antilocalization (WAL) in the system, which demonstrates the presence of spin-orbit coupling (SOC) in γ-GeSe.[34-37] Interestingly, the experimentally measured physical properties of γ-GeSe are analogous to those of GeTe but in stark contrast to those of α-GeSe. Owing to its unique local bonding configuration and crystal symmetry, γ-GeSe serves as a model system for understanding the correlation between the physical properties and polymorphic configurations of group-IV monochalcogenides.

**Results**

**Figure 1** summarizes the crystal structures of Ge-based monochalcogenide polymorphs. An orthorhombic puckered structure shared by α-GeSe and GeS is an archetypal polymorph in which Ge (or Se) atoms form covalent bonds with the three nearest Se (or Ge) atoms. This structure satisfied the 8-*N* rule for covalent bonding.[2] β-GeSe has a local bonding configuration similar to that of α-GeSe, in which the 8-*N* rule is satisfied with a coordination number of three.[24] The main subject of the current study, γ-GeSe, is a recently identified polymorph of GeSe that shows a unique local bonding configuration.[30] In its four-atom-thick layer, the coordination number of Ge is six, bonding with three Ge and three Se atoms. The local coordination of Se is three, and a van der Waals gap exists between Se layers. Thus, the effective coordination number of γ-GeSe can be assigned as 4.5, exceeding the coordination number expected by the 8-*N* rule but less than that of metallic bonding (eight or twelve). α-GeTe (and β-GeTe) shows an effective coordination number of six and did not satisfy the 8-*N* rule.

We measured the electrical transport properties of γ-GeSe using nanofabricated devices with Hall bar electric contact geometry, as shown in **Figure 2**. γ-GeSe flakes were synthesized using a previously reported chemical vapor deposition (CVD) process.[30] The samples usually display a dagger-shape, and the crystal was mainly identified by Raman spectroscopy and transmission electron microscopy (TEM), as shown in **Supporting figure S1**. We fabricated the devices using standard e-beam lithography. Device fabrication with Ti/Au contact typically results in unstable device operation. Under an electrical bias of approximately 1 V, the resistance of the devices began to increase, as shown in **Supporting figure S2**. However, Pt contact led to ultra-high stable device operation even under a bias bigger than 2.5 V. The ambient exposure of γ-GeSe flakes did not have a significant effect on the electrical conductivity, as shown in **Supporting figure S3**, which demonstrates the stability of γ-GeSe.

Pt-contacted γ-GeSe devices were measured at various temperatures (from 2 K to 300 K) under magnetic fields ranging from −9 T to 9 T. The measured electrical conductivity of γ-GeSe was found to be significantly high, approximately $7.4 \times 10^5$ S/m at room temperature and above $1.0 \times 10^6$ S/m at 2 K, as shown in **Figure 2a**. The temperature dependence of the measured resistance indicates that γ-GeSe exhibits a highly-doped or metallic behavior. The resistance increased monotonically in the moderate temperature range (50–300 K), indicating the importance of electron-phonon scattering in this temperature range. From the sign of the measured Hall voltage, we also determined that γ-GeSe is a *p*-type material (**Figure 2b**). The measured Hall voltage shows linearity at all temperatures, suggesting that only hole carriers participate in the electrical transport. The extracted hole concentration was approximately $5 \times 10^{21}$ cm$^{-3}$ and did not show significant temperature dependence (**Figure 2c**). The measured carrier concentration of γ-GeSe was comparable to other highly doped materials.[18,19,38] We note that this experimental result is in contrast to previous first-principles calculations, which showed that bulk γ-GeSe is a semiconductor with a bandgap of approximately 0.4 eV.[33] **Figure 2d** shows that the hole mobility is 12.2 cm$^2$/Vs at room temperature. The temperature dependence of the measured hole mobility indicates that impurity scattering (dominant below 100 K) and phonon scattering (dominant above 150 K) were present in the system.

The thermoelectric properties of γ-GeSe were also investigated. Local thermometers based on four-probe measurements of electrode resistance as well as a microheater were fabricated on a γ-GeSe flake, as shown in **Figure 3a**.[39] Prior to the Seebeck voltage measurements between the two electrodes, thermometer calibration was performed (see **Supporting Figure S4**). The application of Joule heating from the microheater located at the top (or bottom) side of the device induced a temperature gradient and voltage drop across the γ-GeSe flake (**Supporting Figure S5**). Using this procedure, we extracted the Seebeck coefficient ($S = -dV/dT$) of γ-GeSe from 78 K to 470 K (**Figure 3b**). The average Seebeck coefficient at room temperature is approximately 9.4 μV/K, and it increases linearly with temperature. The observed temperature dependence of the Seebeck coefficient is consistent with the Mott formula for heavily doped materials, $S = \frac{8\pi^2 k_B^2}{3eh^2}(\frac{\pi}{3p})^{\frac{2}{3}} m^* T$, where $k_B$ is the Boltzmann constant, $m^*$ is the effective mass, and $p$ is the hole concentration. The theoretically estimated value of the Seebeck coefficient is consistent with the experimental measurements.

**Figure 3c** shows the thermoelectric power factor ($PF = \sigma S^2$) at various temperatures below 500 K. At room temperature, γ-GeSe exhibited a power factor of approximately 66.7 μW/mK$^2$. **Figure 3d** shows a comparison of the room-temperature Seebeck coefficient and power factor of γ-GeSe with those of other group-IV monochalcogenides.[6,18,19,40,41] The thermoelectric power factor of γ-GeSe is comparable to that of other group-IV compounds. Interestingly, the Seebeck coefficient of γ-GeSe is similar to those of GeTe and SnTe but significantly different from those of other Se- or S-based monochalcogenides such as α-GeSe. The measured Seebeck coefficient and power factors of γ-GeSe are smaller compared to recent theoretical calculation results for γ-GeSe[31] because the doping level of γ-GeSe is significantly higher than the values assumed for the calculation.

First-principles calculations based on density functional theory (DFT) were performed to understand the origin of the high *p*-doping mechanisms of γ-GeSe. Detailed calculation methods are provided in the **Supporting Information**. The electronic band structure of monolayer γ-GeSe without defects in a 3 × 3 supercell is shown in **Figure 4a**. The calculations based on DFT confirm that monolayer γ-GeSe is a semiconductor with an indirect bandgap of 0.6 eV, which is consistent with previous calculations.[33] We note that the electronic band structure shows a zone folding effect compared to the band structure calculation based on the primitive unit cell. The calculated bandgap strongly depends on the number of layers in γ-GeSe and the DFT functional used in the calculation methods. In particular, previous calculations indicate that the bandgap of γ-GeSe decreases as the number of layers increases.[33] However, DFT calculations without hybrid functionals usually underestimate the bandgap size compared to the experimental values and the GW calculation.[33] Combining these two effects, we found that the DFT-based bandgap (~0.6 eV) of monolayer γ-GeSe is comparable to the GW calculation-based bandgap (~0.4 eV) of bulk γ-GeSe, which provides a reasonable platform to study the doping effects in bulk γ-GeSe.

We consider vacancy defects in γ-GeSe to be the major source of the doping effect. **Figure 4b and 4c** show the calculated band structures of the 3 × 3 supercell with Se and Ge vacancies, respectively. With the introduction of vacancy defects, band structures undergo significant modifications. Se vacancies introduce extra defect-associated states near the Fermi energy level, as shown in **Figure 4b**. The red lines in the band structure indicate the electronic states associated with the Ge atoms (marked red in the top panel) near a Se vacancy. Newly introduced states are located either above or below the Fermi energy; however, the semiconducting behavior of γ-GeSe is maintained, ruling out Se vacancies from the source of the observed doping effect. However, when Ge vacancies are introduced, the Fermi level is shifted down, reproducing significant *p*-type behavior, as shown in **Figure 4c**. Therefore, the observed *p*-type doping was consistent with the formation of Ge vacancies. At this point, it is worthwhile to perform the spin-unrestricted (spin-polarized) calculations to examine the effect of vacancies on spin splitting or magnetic properties of γ-GeSe. It was found that there are no spin splitting and magnetic moments resulting from the presence of the vacancies, as shown in **Supporting Figure S6**. Therefore, we considered only the spin-restricted case for γ-GeSe throughout this paper.

We considered the thermodynamics of vacancy formation in γ-GeSe. **Figure 4d** shows the vacancy formation energies of γ-GeSe in various chemical environments. The formation energy of Ge vacancies ($E_{v,Ge}$) is significantly smaller than that of Se vacancies ($E_{v,Se}$) in all chemical environments. Owing to the low vaporization temperature of Se, the chemical environment during γ-GeSe synthesis is likely to be Se-rich, which further lowers the formation energy of the Ge vacancies. **Supporting Figure S7a** compares the vacancy formation energies in Ge monochalcogenides from the literatures.[42-44] Overall, the formation energy of Ge vacancies was smaller than that of chalcogen vacancies for all monochalcogenides. Moreover, as shown in **Supporting Figure S7a**, the significantly lower formation energy of Ge vacancies is attributed to both γ-GeSe and GeTe, which provides strong evidence that the highly *p*-doped behavior shared by γ-GeSe and GeTe can be attributed to Ge vacancies. However, α-GeSe and GeS show only a moderate level of doping, which is also consistent with our interpretation. **Supporting Figure S7b** displays the electrical conductivity and carrier concentration of some Ge monochalcogenides.[18-20,24] Materials with a low Ge vacancy formation energy exhibit high carrier concentration, resulting in high electrical conductivity.

Elemental analysis based on energy dispersive spectroscopy (EDS) was utilized to experimentally verify the formation of Ge vacancy in γ-GeSe. To precisely determine the Ge:Se ratio in γ-GeSe (**Figure 4e**), we used α-GeSe as a reference sample because the stoichiometric Ge:Se ratio in α-GeSe is very close to 1:1 due to its low doping level. By EDS elemental analysis, we indeed confirmed that γ-GeSe is Ge-deficient compared to α-GeSe and the estimated Ge vacancy concentration in γ-GeSe is approximately 5% as shown in **Figure 4f** and **Supporting Figure S8**. The experimentally measured Ge vacancy concentration is consistent with the observed carrier concentration.

To better understand the mechanism of forming Ge vacancies in γ-GeSe, we have performed the projected crystal orbital Hamiltonian population (-pCOHP) analysis.[45] **Figure 4g-i** shows the -pCOHP curves of the covalent Ge-Ge and Ge-Se interactions in the case of perfect γ-GeSe, γ-GeSe structure with Se vacancy, and Ge vacancy, respectively. The positive/negative values in the -pCOHP curves represent the bonding/antibonding states, and the magnitude indicates the strength of the bonding/antibonding interactions. The relatively strong Ge–Se antibonding interactions are shown near the valence band maximum of γ-GeSe. It is found that the band states of the cationic Ge atoms are located near the Fermi level. In contrast, most band states of the anionic Se atoms are formed at the deep level of the valence band, so removing Ge atoms from the γ-GeSe is energetically more favorable than Se atoms. Upon the removal of Ge atoms, the antibonding interaction region below the Fermi level moves upward, making the occupied antibonding states reduced. Therefore, the introduction of Ge vacancy provides a way for the stabilization of in γ-GeSe. We note that a recently published paper performed a similar analysis and confirmed that Ge vacancy formation can reduce the antibonding state in γ-GeSe.[46]

Next, we discuss the effect of SOC on γ-GeSe, as revealed by magnetoresistance (MR). The MR results showed a typical quadratic dependence on the magnetic field at all temperatures, as shown in **Figure 5a**. Moreover, the cusp of the MR graph was clearly observed near the zero field, as shown in the inset of **Figure 5a**. The observed cusp of the MR graph at low temperatures (< 20 K) is due to the WAL effect, which is a typical signature of the existence of SOC. The interference between the time-reversible paths inside $l_\varphi$ (phase coherence length) leads to the WAL effect, and the application of the magnetic fields breaks this coherence. This observation is consistent with a previous theoretical calculation, which suggested the existence of SOC in γ-GeSe.[33] **Figure 5b** shows Kohler's plot, which is the MR measurement as a function of $B/\rho_{xx}(0)$. Kohler's plot collapses to the same curve for all temperatures if only a single type of carrier contributes to the transport.[35] It appears that Kohler's rule is slightly violated for γ-GeSe, which may originate from the multiple scattering mechanisms of charge carriers.[34]

The MR modulation by WAL is analyzed by the Hikami-Larkin-Nagaoka (HLN) theory[47] under a strong SOC limit:

$$\Delta\sigma(B) = \sigma_{xx}(B) - \sigma_{xx}(0) = -\alpha \frac{e^2}{2\pi^2 \hbar} \left\{ \Psi\left(\frac{1}{2} + \frac{\hbar}{4eBl_\varphi^2}\right) - \frac{1}{2}\ln\left(\frac{\hbar}{4eBl_\varphi^2}\right) \right\}$$

where $\alpha = n/2$ accounts for $n$ conduction channels in two dimensions, $e$ is the elementary charge, $\hbar$ is the reduced Planck constant, and $\Psi$ is the digamma function. The fitting based on HLN theory shows a reasonable match with the experimental measurements, as shown in **Figure 5c**. **Figure 5d** shows the temperature-dependent behavior of the fitting constants $l_\varphi$ and $\alpha$. $\alpha$ is estimated to be slightly larger than 0.5 and is nearly constant below 20 K. The temperature dependence of the extracted $l_\varphi$ exhibits $T^{-0.67}$ behavior, which shows a faster decay than that of Nyquist electron-electron scattering.[48] This deviation of the fitting parameters from 2D transport can be attributed to bulk-type conduction in γ-GeSe owing to the thickness of the measurement samples (~ 80 nm).[49] A similar WAL was also reported for GeTe,[36] which is another similarity between γ-GeSe and GeTe.

**Conclusion**

In this study, the temperature-dependent electrical and thermoelectric properties of γ-GeSe were experimentally established for the first time. From the Hall measurements and the measured Seebeck coefficient, we found that the γ-GeSe is a *p*-doped material with a high carrier concentration (~ $5 \times 10^{21}$ cm$^{-3}$). Through elemental analysis and first-principles calculations, we confirmed that the main source of the observed high *p*-doping is Ge vacancy in the new polymorphic configuration. Furthermore, the WAL and SOC were confirmed by MR measurements of γ-GeSe. The local bonding configuration and measured physical properties of γ-GeSe share similarity to those of GeTe but are in stark contrast to those of α-GeSe. Considering that GeTe is often categorized as having remarkable electronic state dubbed metavalent bonding with phase-change memory applications,[16,50,51] we believe that the electronic state in γ-GeSe also possesses a unique nature that is highly suitable for related applications. Finally, recent theoretical studies demonstrate that the thermoelectric performance of γ-GeSe with an appropriate carrier concentration is highly favorable mainly due to its low lattice thermal conductivity (**Supporting Table 1**). Therefore, we envision that the doping control of γ-GeSe, particularly the reduction of carrier concentrations, will greatly increase thermoelectric performance and open up a new avenue for thermoelectric applications of γ-GeSe, similar to the case of GeTe.[7,52]

**Notes**

The authors declare no competing financial interest.

**Acknowledgements**

This work was supported by the Basic Science Research Program of the National Research Foundation of Korea (NRF-2017R1A5A1014862, NRF-2021R1A4A1031900, NRF-2022R1A2C4002559, NRF-2022R1A2C1006740, NRF-2022R1I1A1A01070953, and NRF-2022R1A6A3A13073662), Yonsei Signature Research Cluster Program of 2022 (2022-22-0004), and the Institute for Basic Science (IBS-R026-D1). This work was also supported by Global Research and Development Center Program (2018K1A4A3A01064272), the Engineering Research Center (ERC) program (NRF- 2019R1A5A1027055), and Samsung Electronics Co., Ltd (IO210202-08367-01)

**Supporting Information Available:** Method section, extra characterizations of γ-GeSe, additional experimental data including the stability under ambient exposure, thermoelectric measurements of γ-GeSe, EDS elemental analysis, and extra first principles calculation results.

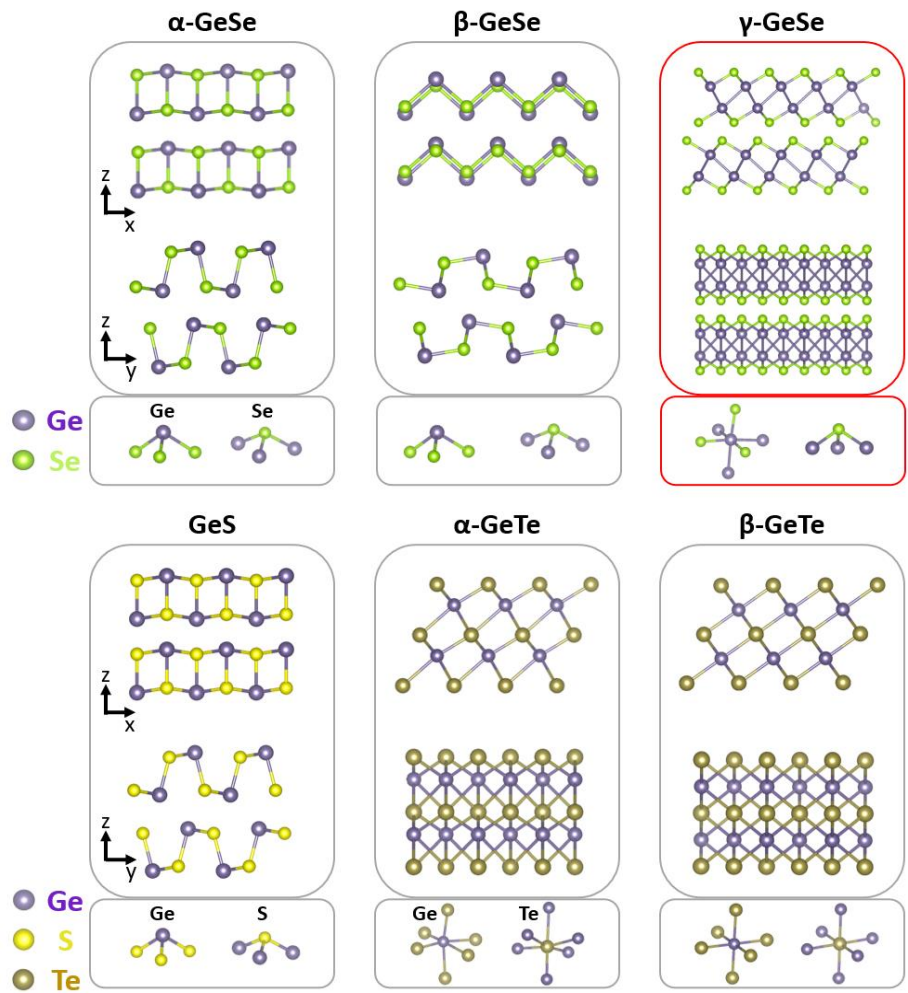

**Figure 1. Crystal structures and local bonding configurations of Ge monochalcogenides.** For each panel, the top schematics show side views of the crystal from two directions, and the bottom schematics show the local bonding configurations centered at Ge and Se (or S/Te) atoms.

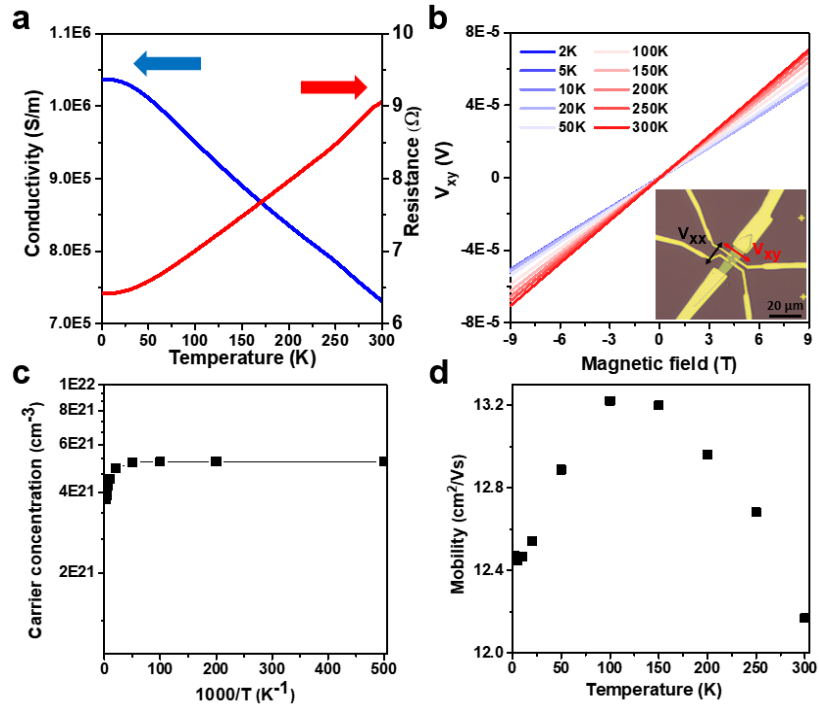

**Figure 2. Electrical transport properties of γ-GeSe.** (a) Temperature-dependent resistance and electrical conductivity. (b) Temperature-dependent Hall voltage as a function of applied magnetic field. The inset shows the optical image of a typical device for transport measurement. (c) Charge carrier concentration as a function of temperature. (d) Temperature-dependent charge carrier mobility based on Hall measurement.

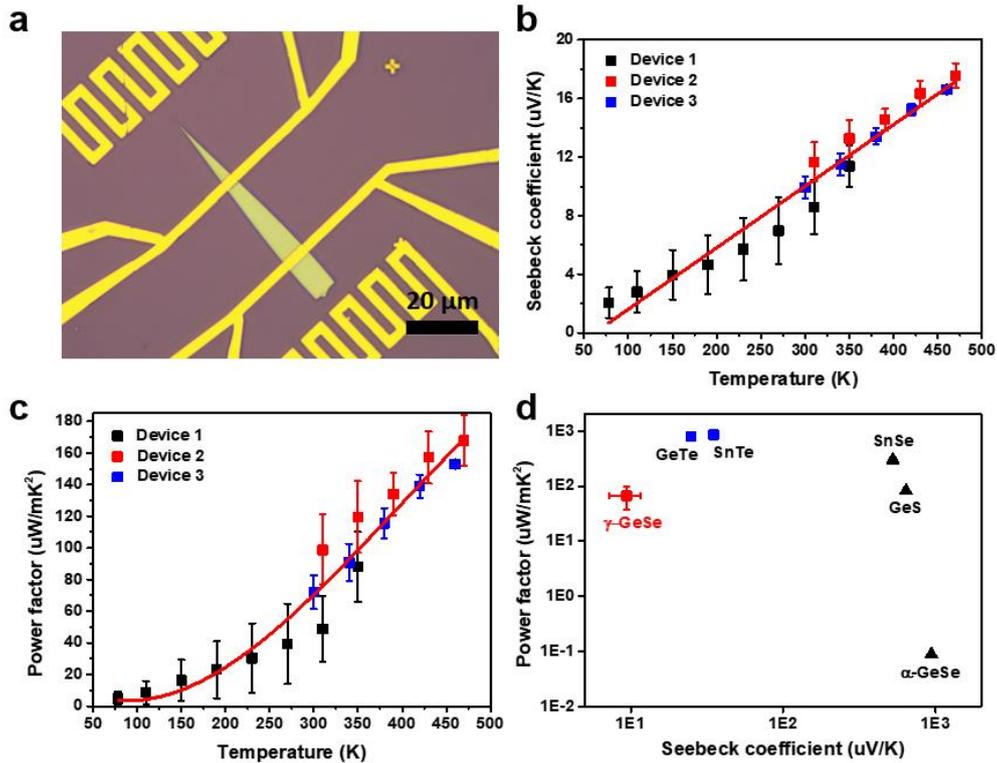

**Figure 3. Thermoelectric measurements of γ-GeSe.** (a) Optical image of a typical device for thermoelectric measurements. (b) Measured Seebeck coefficient as a function of temperature. The data point shown in different colors indicate measurements from different devices. The red line is a linear fitting to the data. (c) Thermoelectric power factor as a function of temperature. The data point shown in different colors indicate measurements from different devices. The red curve is a fitting to the data by cubic-temperature dependence. (d) Comparison of Seebeck coefficients and power factors among selected group-IV monochalcogenides.

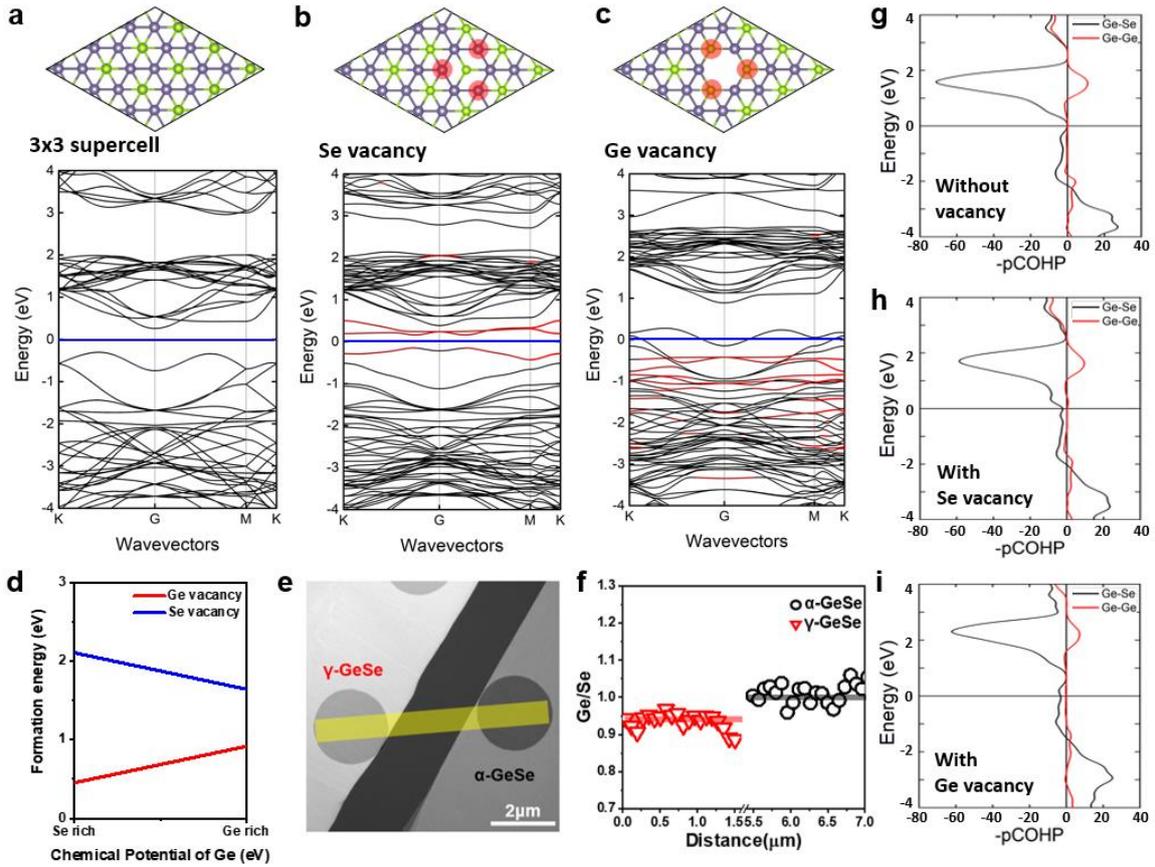

**Figure 4. First-principles calculations for p-doping mechanism.** (a) Upper: 3×3 supercell of monolayer γ-GeSe, low: electrical band structures of 3×3 supercell. The blue horizontal line indicates the Fermi energy of the system. (b) Electrical band structure of 3×3 supercell with a Se vacancy. The red lines in the band structures correspond to the states from Ge atoms near the Se vacancy (marked with red color in the upper panel). (c) Electrical band structure of 3×3 supercell with a Ge vacancy. The red lines correspond to the states from marked Se atoms near the Ge vacancy. (d) Calculated formation energies for Ge and Se vacancies in γ-GeSe. (e) STEM image of α-GeSe and γ-GeSe flakes transferred onto a holey $Si_3N_4$ TEM grid. The marked linear area is used for Ge/Se ratio profile in panel f. (f) Ge/Se ratio profile from γ-GeSe and α-GeSe. (g-i) COHP bonding/antibonding analysis curves in the case of perfect γ-GeSe, γ-GeSe structure with Se vacancy, and Ge vacancy, respectively. The black and red solid lines represent Ge-Se and Ge-Ge bonding/antibonding interactions, respectively. Bonding interactions to the right, antibonding interactions to the left.

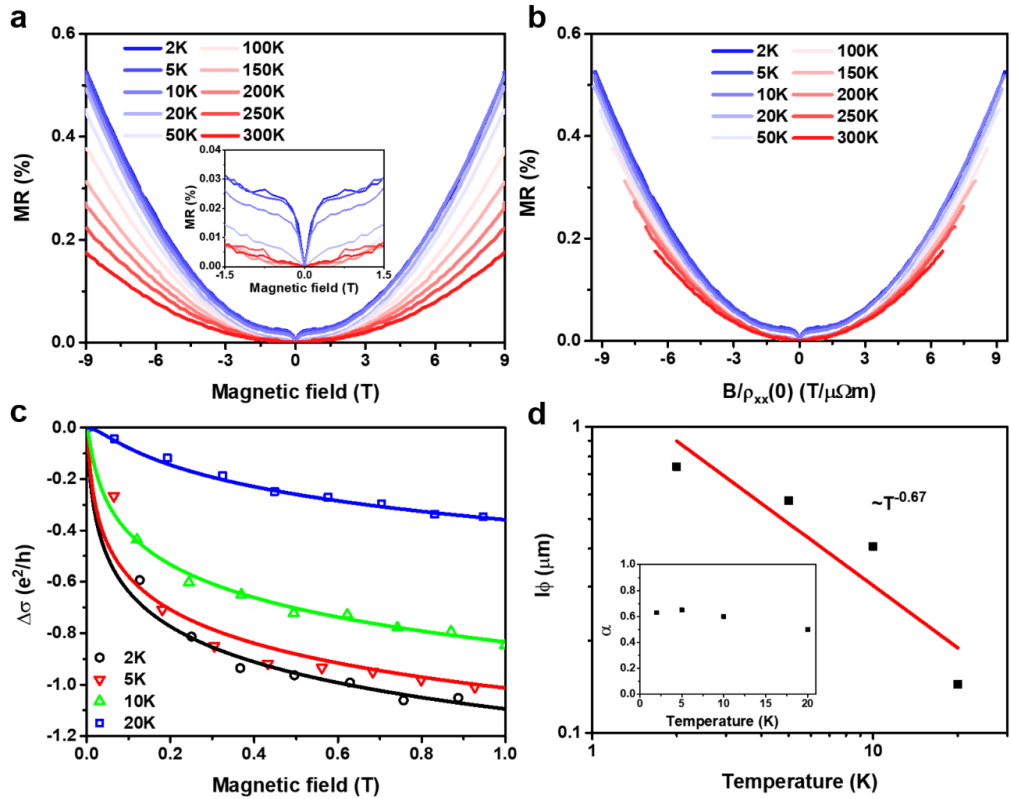

**Figure 5. MR measurements of γ-GeSe and analysis.** (a) MR measurements at various temperatures. The inset shows the zoom-in data near zero magnetic field. (b) Kohler's plot at various temperatures. (c) WAL effect analysis with fitting based on HLN theory. (d) Temperature dependence of fitting parameters.

**Table of Contents graphic**

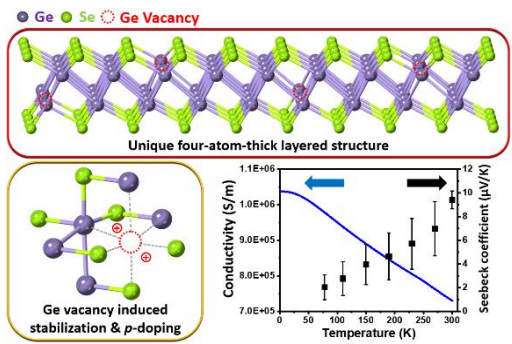